\documentclass[10pt,aps,prd,twocolumn,superscriptaddress,floatfix,amsmath,amssymb,amsfonts,longbibliography,nofootinbib]{revtex4-2}

\usepackage{placeins}

\usepackage{comment}
\usepackage[normalem]{ulem}
\usepackage[english]{babel}
\usepackage{graphicx}% Include figure files
\usepackage{dcolumn}% Align table columns on decimal point
\usepackage{bm}% bold math
\usepackage{blindtext}
\usepackage{verbatim}
\usepackage{mathrsfs}
\usepackage{musicography}
\usepackage{amsmath}
\usepackage{dsfont}
\usepackage{mathrsfs}
\usepackage{amssymb}
\usepackage{amsthm}
\usepackage{cancel}
\usepackage{physics}
\usepackage{epstopdf}
\usepackage{mathtools}
\usepackage{color}
\usepackage{epsfig}
\usepackage{pst-grad} % For gradients
\usepackage{pst-plot} % For axes
\usepackage{hyperref}
\usepackage{verbatim}
\usepackage{afterpage}
\usepackage{sidecap}
\usepackage{booktabs}

\usepackage{adjustbox}
\hypersetup{
    colorlinks=true,
    linkcolor=blue,
    filecolor=red,      
    urlcolor=cyan,
}
\usepackage{tikzsymbols}
\begin{document}
%\title{Neutrino Decoherence in \texorpdfstring{\(\kappa\)}{kappa}-Minkowski Quantum Spacetime: An Open Quantum Systems Paradigm}
\title{Stochastic Quantization of Electrodynamics and Linearized Gravity}
\author{Partha Nandi}
\email{pnandi@sun.ac.za}
\affiliation{Department of Physics, University of Stellenbosch, Stellenbosch, 7600, South Africa}
\affiliation{National Institute of Theoretical and Computational Sciences (NITheCS), Stellenbosch, 7604, South Africa}

\author{Partha Ghose}
\email{partha.ghose@gmail.com}
\affiliation{Tagore Centre for Natural Sciences and Philosophy,\\ Rabindra Tirtha, New Town, Kolkata 700156, India}

%\author{Partha Ghose}
%\email{partha.ghose@gmail.com}
%\affiliation{Tagore Centre for Natural Sciences and Philosophy,\\ Rabindra Tirtha, New Town, Kolkata 700156, India}

%\author{A. S. Majumdar}
%\email{archan@bose.res.in}
%\affiliation{S. N. Bose National Centre for Basic Sciences, Salt Lake, Kolkata 700106, India}

%\author{Graeme Pleasance}
%\email{gpleasance1@gmail.com}
%\affiliation{Department of Physics, University of Stellenbosch, Stellenbosch, 7600, South Africa}
%\affiliation{National Institute of Theoretical and Computational Sciences (NITheCS), Stellenbosch, 7604, South Africa}

%\author{Francesco Petruccione}
%\email{petruccione@sun.ac.za}
%\affiliation{Department of Physics, University of Stellenbosch, Stellenbosch, 7600, South Africa}
%\affiliation{National Institute of Theoretical and Computational Sciences (NITheCS), Stellenbosch, 7604, South Africa}
%\affiliation{School of Data Science and Computational Thinking, University of Stellenbosch, Stellenbosch 7600, South Africa}

%{\bf Keywords}: stochastic mechanics, electrodynamics, Poisson process, photon, Linearized gravity, graviton, 
\begin{abstract}

We develop a unified stochastic framework in which a velocity- and helicity-reversing Poisson process gives rise to the Telegrapher’s equation. Analytic continuation to the complex plane results in Dirac-like evolution equations for electromagnetic and linearized gravitational fields. A small but nonzero mass parameter is essential to enable helicity reversals. Yet, the correct massless wave equations are recovered as the physically relevant massless limit is approached smoothly, with the singular point $m = 0$ excluded from the construction. Remarkably, probability does not enter as an external postulate—as in the Born rule in standard quantum mechanics —but is intrinsic to the stochastic process. This probabilistic structure becomes embedded in the wave fields through a natural rescaling by the Planck length.

\end{abstract}
\maketitle
\section{Introduction}

By now, it is widely recognized that perturbative approaches to quantum gravity are fundamentally inconsistent~\cite{Ashtekar198}. The challenge of reconciling general relativity and quantum mechanics—commonly known as the problem of quantum gravity—remains one of the most profound open questions in theoretical physics~\cite{1,2,Sugiyama2024}. General relativity describes gravity as the curvature of spacetime and governs large-scale phenomena such as black holes and gravitational waves~\cite{2+}, while quantum mechanics governs the microscopic world through probabilistic evolution and wavefunction collapse during measurement~\cite{3,Pittaway:2021yby}. These two frameworks rest on fundamentally distinct principles and are difficult to unify within a common theoretical structure.

At extremely short distances and timescales approaching the Planck length, both theories are expected to break down in their standard formulations. Neither general relativity nor quantum mechanics, taken individually, provides a consistent description under such extreme conditions~\cite{Nandi:2022noj,Nandi:2023tfq}. This inconsistency points to the need for a new, unified framework—quantum gravity—that can reconcile their foundational differences. However, despite decades of effort, no direct experimental evidence has yet revealed the quantum nature of gravity~\cite{4,5}. The absence of empirical guidance makes the search for quantum gravity particularly challenging and leaves major open problems unresolved, including the nature of dark matter, dark energy, and the cosmological constant problem~\cite{6}.

Although string theory remains a leading candidate~\cite{7}, it faces ongoing conceptual and technical challenges~\cite{6,8}. In parallel, recent proposals have emerged for probing quantum features of gravity indirectly, such as through graviton-induced fluctuations in gravitational wave detectors~\cite{PhysRevD.103.044017}, or entanglement generation via gravitational interactions, as explored in~\cite{2}.

This situation prompts a broader question: could the tension between general relativity and quantum mechanics stem not only from gravity, but also from the foundations of quantum theory itself? In particular, the transition from microscopic quantum behavior to classical determinism remains poorly understood. Penrose has suggested that gravity may play a role in quantum state reduction~\cite{5}, proposing that spacetime structure could be tied to wavefunction collapse~\cite{8+}.

From a conceptual standpoint, two major approaches to quantizing gravity are often discussed: one treats gravity as a quantum field on a classical background, while the other seeks to quantize spacetime geometry itself~\cite{9,10}. While these approaches are deeply interconnected, neither guarantees that gravity must be quantized. In fact, quantizing linearized gravity leads to a nonrenormalizable quantum field theory. Nevertheless, effective field theory techniques have shown that meaningful low-energy predictions can still be extracted~\cite{11}. Early work by Feynman and others suggested that general relativity might emerge as a classical limit of a deeper quantum framework.

An alternative approach revisits the foundations of quantum mechanics. Recent insights~\cite{11+} suggest that stochastic processes—such as Wiener processes for continuous motion or Poisson processes for discrete jumps—could underlie the statistical structure of quantum theory. In this perspective, quantum mechanics may emerge from a more fundamental stochastic dynamics~\cite{Breuer1999}. Nelson’s stochastic quantization, following earlier ideas by Fényes and Kershaw, showed that the Schr\"{o}dinger equation could be derived from an underlying Brownian motion~\cite{nel1,nel2,nel3}. More recently, extensions using Poisson processes have recovered relativistic quantum equations, including the Dirac equation~\cite{gav}.

Motivated by these developments, we construct a unified stochastic quantization framework for both electromagnetic and linearized gravitational fields. Although these differ in gauge structure and physical content, we show that both admit a formally analogous treatment in the weak-field limit via Poisson-based stochastic processes. Our aim is to demonstrate that wavefunctions for both photons and gravitons can emerge from a time-symmetric stochastic dynamics analogous to Nelson’s construction.

In the electromagnetic case, we work directly with gauge-invariant electric and magnetic fields, constructing a Schr\"{o}dinger-like equation using the Riemann–Silberstein vector. This yields a helicity-resolved description of the photon field in a manifestly gauge-invariant form, and provides a clear probabilistic interpretation based on stochastic helicity-flip dynamics.

In the gravitational case, the situation is more subtle. Since linearized gravity involves the gauge field \( h_{\mu\nu} \), isolating the physical degrees of freedom requires gauge fixing. We adopt the transverse-traceless (TT) gauge, which projects onto the two propagating helicity-\( \pm 2 \) modes. This enables us to define a graviton wavefunction analogous to the photon case.

Unlike Nelson’s scalar-based stochastic mechanics, our framework begins with the classical field equations of electrodynamics and linearized gravity. We reformulate these as Schr\"{o}dinger-like equations for helicity-resolved field amplitudes and derive them from a Poisson-type stochastic process. Importantly, we work entirely with physical, gauge-invariant field components—Riemann–Silberstein vectors for photons and TT-projected metric perturbations for gravitons—without invoking field operators or canonical quantization. This yields a realist, probabilistic interpretation of wave propagation rooted in classical randomness.

Our construction introduces a small but nonzero mass parameter that enables helicity-reversing stochastic processes. This mass acts as a regulator, allowing coherent, wave-like behavior to emerge from the underlying dynamics. The physically relevant massless limit is then approached smoothly, with the singular point \( m = 0 \) excluded from the stochastic formulation—yet the correct massless wave equations are recovered without discontinuity. We also perform a rescaling at the Planck length, providing the correct dimensional foundation for interpreting amplitude-squared quantities as statistical weights. Notably, probability does not enter as an external postulate—as in the Born rule—but emerges dynamically from the stochastic evolution itself.

The main aim of this work is to provide a first-principles derivation of quantum-like wave equations for massless helicity-1 and helicity-2 fields—photons and gravitons—based on a stochastic process governed by discrete Poisson dynamics. By reformulating classical field equations into Schr\"{o}dinger-type evolution equations for helicity-resolved amplitudes, we show that both electrodynamics and linearized gravity admit a common probabilistic foundation. Crucially, this framework avoids canonical quantization and instead recovers quantum behavior as the emergent limit of a time-symmetric stochastic process, with probability arising naturally rather than being postulated. This provides a unified, gauge-invariant, and realist interpretation of wave propagation in massless gauge theories, rooted in classical randomness and Planck-scale structure.

A conceptually complementary approach has recently been developed by Przanowski \textit{et al.}~\cite{Przanowski:2024}, who constructed a quantum wavefunction for the linear graviton using the Weyl spinor formalism. By reformulating the Bianchi identities in terms of totally symmetric spinors, they derived Schr\"{o}dinger-like evolution equations for the helicity \( \pm 2 \) components of the gravitational field. While their formulation is rooted in canonical quantization, both approaches share a key structural feature: graviton propagation is captured by first-order helicity-resolved dynamics. Our results thus offer a stochastic foundation that parallels the spinorial quantum graviton picture, providing an alternative route to graviton wavefunctions without quantizing the gravitational field.

\vspace{0.3cm}
\noindent

The remainder of this paper is organized as follows. Section~II reviews Nelson’s stochastic mechanics and its extensions to relativistic systems. In Section~III, we apply this framework to electrodynamics, constructing the photon wavefunction and deriving a Schr\"{o}dinger-like equation from an underlying Poisson process. Section~IV extends the approach to linearized gravity, formulating the graviton wavefunction in the transverse-traceless (TT) gauge. In Section~V, we develop a stochastic quantization of the graviton field and explore its physical implications. Sections~VI and~VII summarize our main results and outline possible directions for future research. Section~VIII provides an appendix, using the plane wave method to identify the physical degrees of freedom in linearized gravity.

\section{A Brief Review of Stochastic Mechanics}

In this section, we present a brief overview of stochastic mechanics, highlighting the essential concepts relevant for the stochastic quantization of electrodynamics and weak gravity discussed later.

Stochastic mechanics was first developed by Nelson~\cite{nel1,nel2}, who demonstrated that quantum mechanics can arise from an underlying classical stochastic process, rather than being imposed axiomatically. In this framework, quantum randomness is attributed to the continuous but nowhere differentiable Brownian motion of particles within a hypothetical ``ether.''

%\textbf{Stochastic mechanics was first developed by Nelson~\cite{nel1,nel2}, who has shown that quantum mechanics can emerge from an underlying classical stochastic process, rather than being postulated a priori. In this framework, the randomness of quantum behavior is attributed to a continuous yet nowhere differentiable Brownian motion of particles within a hypothetical ``ether.''}

A key aspect of Nelson's approach is the description of particle trajectories using two diffusion processes—one evolving forward in time and the other backward. These forward and backward stochastic evolutions are governed by distinct but related probability distributions. Their interplay naturally leads to the emergence of quantum mechanical features, including the Schr\"{o}dinger equation.

%In this section, we provide a concise review of stochastic mechanics, emphasizing the key ideas that will be instrumental for the later stochastic quantization of electrodynamics and weak gravity. 

%Stochastic mechanics, originally formulated by Nelson~\cite{nel1,nel2}, proposes that quantum mechanics can be derived from an underlying classical stochastic process rather than postulated independently. In this framework, the indeterminism of quantum phenomena arises from a continuous but nowhere differentiable Brownian motion of particles immersed in a hypothetical ``ether.'' 

%A central feature of Nelson's approach is the modeling of particle trajectories using a pair of diffusion processes: one evolving forward in time and one evolving backward. These two stochastic evolutions are governed by distinct, though related, probability densities. The interplay between the forward and backward processes leads naturally to the emergence of quantum mechanical structures, such as the Schr\"{o}dinger equation.

\subsection{Nonrelativistic Stochastic Mechanics}
\label{subsec:nonrel_stochastic}

Nelson’s stochastic mechanics is built upon four core physical postulates, from which the formal structure of quantum mechanics—particularly the Schr\"{o}dinger equation—can be derived. The essential idea is to reinterpret quantum dynamics as emerging from an underlying stochastic process, akin to Brownian motion, but with diffusion strength set by Planck's constant.

\begin{enumerate}
\item[(A)] \textbf{Universal Brownian Motion:}  
Particles experience a universal Brownian motion, even in the vacuum, modeled as an ether-like medium. This stochastic influence is frictionless, ensuring Galilean invariance, and conserves energy in an average sense. The underlying motion is described by a Markov process in configuration space, resulting in conservative diffusion: local energy fluctuations exist, but average out to zero over time.

\item[(B)] \textbf{Stochastic Differential Equations (SDEs):}  
In one spatial dimension, a stochastic process \( X(t) \) is governed by the Itô stochastic differential equation:
\begin{equation}
dX(t) = b(X(t),t)\,dt + \sigma\,dW(t),
\end{equation}
where \( b(X(t),t) \) is the drift coefficient evaluated along the stochastic trajectory, \( \sigma \) is the noise amplitude (with dimensions of velocity), and \( W(t) \) is a standard Wiener process. This formulation describes how a particle’s trajectory evolves under both deterministic drift and random fluctuations.

The corresponding ensemble evolution of the probability density \( \rho(x,t) \) is governed by the Fokker–Planck equation:
\begin{equation}
\frac{\partial \rho}{\partial t} = -\frac{\partial}{\partial x} \big(b(x,t) \rho\big) + \frac{1}{2} \sigma^2 \frac{\partial^2 \rho}{\partial x^2}.
\label{fp_general}
\end{equation}
Here, \( x \) denotes a fixed spatial coordinate, and the function \( b(x,t) \) represents the same drift field as in the SDE, but evaluated over the deterministic variable \( x \). The term \( \frac{1}{2} \sigma^2 \) is referred to as the diffusion coefficient and characterizes the rate at which probability spreads spatially due to stochastic fluctuations.

Nelson's stochastic mechanics postulates that two such SDEs exist: one forward in time and one backward, each with identical diffusion strength:
\begin{align}
dX(t) &= b_f(X(t),t)\,dt + \sigma\,dW_f(t), \\
dX(t) &= b_b(X(t),t)\,dt + \sigma\,dW_b(t),
\end{align}
where \( b_f \) and \( b_b \) denote the forward and backward drift fields, and \( W_f(t) \), \( W_b(t) \) are independent standard Wiener processes.

The corresponding forward and backward Fokker–Planck equations for the probability density are:
\begin{align}
\frac{\partial \rho}{\partial t} &= -\frac{\partial}{\partial x}(b_f(x,t)\, \rho) + \frac{1}{2} \sigma^2 \frac{\partial^2 \rho}{\partial x^2}, \\
\frac{\partial \rho}{\partial t} &= -\frac{\partial}{\partial x}(b_b(x,t)\, \rho) - \frac{1}{2} \sigma^2 \frac{\partial^2 \rho}{\partial x^2}.
\end{align}
This pair of equations reflects the time-symmetric nature of Nelson's framework and plays a central role in deriving the quantum mechanical continuity and Hamilton–Jacobi equations from the underlying stochastic process.

\item[(C)] \textbf{Diffusion Coefficient from Planck’s Constant:}  
Nelson identifies the diffusion coefficient with quantum scales:
\begin{equation}
\sigma^2 = \frac{\hbar}{m},
\end{equation}
so that quantum effects emerge from stochastic fluctuations. The equations above become:
\begin{align}
\frac{\partial \rho}{\partial t} &= -\frac{\partial}{\partial x}(b_f \rho) + \frac{\hbar}{2m} \frac{\partial^2 \rho}{\partial x^2}, \\
\frac{\partial \rho}{\partial t} &= -\frac{\partial}{\partial x}(b_b \rho) - \frac{\hbar}{2m} \frac{\partial^2 \rho}{\partial x^2}.
\end{align}

\item[(D)] \textbf{Stochastic Acceleration and Newton’s Law:}  
Because paths $X(t)$ are continuous but nowhere differentiable, Nelson defines generalized velocities via the mean forward and backward derivatives:
\begin{align}
D_f X(t) &= \lim_{\Delta t \to 0^+} E_t\left[\frac{X(t+\Delta t) - X(t)}{\Delta t}\right], \\
D_b X(t) &= \lim_{\Delta t \to 0^+} E_t\left[\frac{X(t) - X(t-\Delta t)}{\Delta t}\right],
\end{align}
where $E_t[\cdot]$ is the conditional expectation given $X(t) = x$. For differentiable trajectories, these reduce to ordinary velocities. Nelson postulates that Newton’s second law remains valid in the stochastic framework if acceleration is defined via the symmetric second-order operator:
\begin{equation}
m\,a(X(t)) := m\frac{1}{2}(D_f D_b + D_b D_f) X(t) = F(X(t)),
\end{equation}
where $F$ is the applied force. This leads to dynamics consistent with both stochasticity and classical mechanics in the appropriate limit.
\end{enumerate}

The forward and backward derivatives yield:
\begin{equation}
D_f X(t) = b_f(X(t),t), \quad D_b X(t) = b_b(X(t),t),
\end{equation}
giving a complete dynamical description in analogy with deterministic velocity fields.

The forward and backward drifts define the \textit{current velocity} $v$ and the \textit{osmotic velocity} $u$::
\begin{align}
v &= \frac{b_f + b_b}{2}, \\
u &= \frac{b_f - b_b}{2}.
\end{align}
allows us to decompose the dynamics as follows:
\begin{itemize}
\item Adding the two Fokker–Planck equations yields the continuity equation:
\begin{equation}
\frac{\partial \rho}{\partial t} + \frac{\partial}{\partial x}(v \rho) = 0,
\label{cont_eq}
\end{equation}
expressing probability conservation.
\item Subtracting them leads to:
\begin{equation}
u = \frac{\hbar}{2m} \frac{\partial_x \rho}{\rho} = \frac{\hbar}{m} \partial_x R,
\label{osmotic}
\end{equation}
where we define $\rho(x,t) = e^{2R(x,t)}$. This identifies $u$ as a gradient flow driven by the probability density.
\end{itemize}

Consequently, the stochastic dynamics for $X(t)$ can be recast in terms of $v$ and $u$:
\begin{align}
dX(t) &= (v + u)\,dt + \sigma\,dW_f(t), \\
dX(t) &= (v - u)\,dt + \sigma\,dW_b(t),
\end{align}
where $\sigma = \hbar/m$ links quantum effects directly to noise intensity.

The current velocity $v$ is curl-free and thus expressible as a gradient:
\begin{equation}
v(x,t) = \frac{1}{m} \partial_x S(x,t),
\label{current_velocity}
\end{equation}
where $S(x,t)$ plays the role of the classical action.

To complete the analogy with mechanics, we introduce a stochastic Lagrangian density:
\begin{equation}
\mathcal{L}(x,t) = \frac{1}{2}m (v^2 - u^2) - V(x),
\end{equation}
and apply a stochastic variational principle, as developed by Guerra and Morato~\cite{gu}, to derive the equations of motion. This yields the Schr\"{o}dinger equation, with the emergence of quantum dynamics encoded in the interplay between $v$, $u$, and the diffusion constant $\sigma = \sqrt{\hbar/m}$.\\

%the action $S$ can be obtained through a stochastic variational principle, as shown by Guerra and Morato~\cite{gu}.

Following the stochastic variational approach, one derives coupled differential equations for $S$ and $R$:
\begin{align}
\frac{\partial S}{\partial t} + \frac{1}{2m} \left( \frac{\partial S}{\partial x} \right)^2 + V + V_Q &= 0; \quad\\
V_Q = -\frac{\hbar^2}{2m} \left( (\partial_x R)^2 + \partial_x^2 R \right),
\label{hjb} \\
\frac{\partial R}{\partial t} + \frac{1}{2m}\left( R \partial_x^2 S + 2 \partial_x R \partial_x S \right) &= 0.
\label{continuity}
\end{align}

Here, $V_Q$ is the so-called \textit{quantum potential}, expressible in terms of $\rho$:
\begin{equation}
V_Q = -\frac{\hbar^2}{4m} \left( \frac{\partial_x^2 \rho}{\rho} - \frac{(\partial_x \rho)^2}{2\rho^2} \right).
\end{equation}

Equations~\eqref{hjb} and~\eqref{continuity} together provide a complete description of the stochastic dynamics.  
Introducing the complex wavefunction
\begin{equation}
\psi(x,t) = \sqrt{\rho(x,t)} \, e^{i S(x,t)/\hbar},
\end{equation}
one finds that these two equations combine to yield the Schr\"{o}dinger equation:
\begin{equation}
i \hbar \frac{\partial \psi}{\partial t} = \left( -\frac{\hbar^2}{2m} \partial_x^2 + V(x) \right) \psi(x,t).
\end{equation}

Thus, quantum mechanics emerges naturally from an underlying classical stochastic process, without invoking the conventional postulates of wavefunction collapse or intrinsic indeterminism.

In stochastic mechanics, the correspondence between the stochastic variables and the quantum wavefunction iscalled ``Nelson map'':
\begin{align}
\rho(x,t) &= |\psi(x,t)|^2, \\
u(x,t) &= \frac{\hbar}{m} \partial_x \Re \ln \psi(x,t), \\
v(x,t) &= \frac{\hbar}{m} \partial_x \Im \ln \psi(x,t),
\label{map1}
\end{align}
where $\rho$ denotes the probability density, $u$ the osmotic velocity, and $v$ the current velocity.  
Once the wavefunction $\psi(x,t)$ is known, the complete stochastic process, including the distribution of trajectories and their drift characteristics, is fully determined.

Stochastic mechanics, particularly in the formulation by Yasue~\cite{yas} and the stochastic control approach developed by Guerra and Morato~\cite{gu}, offer an objective and realist description of microphysical phenomena by modeling quantum behavior as emerging from underlying Brownian motion. It also reveals deep connections to Feynman's path integral formulation~\cite{com, wang}, providing a conceptual bridge between classical stochastic processes and quantum theory. Moreover, the interpretational difficulties arising from the quantum measurement postulate are largely avoided—provided one accepts the physical reality of particle trajectories, as in stochastic mechanics~\cite{gold}.

%In the following sections, we extend these ideas to field-theoretic systems. We demonstrate how the stochastic quantization framework can be systematically applied to photons and gravitons, thereby establishing a unified stochastic description for both quantum electrodynamics and linearized gravity.

\subsection{Comparison with usual Quantum Mechanics}
%\subsection{Comparison with Quantum Mechanics}
\label{subsec:comparison_qm}

There are two distinct probabilistic interpretations of the Schr\"{o}dinger equation:  
(a) the standard quantum mechanical interpretation introduced by Born, and  
(b) the stochastic mechanical interpretation proposed by Nelson.  
This naturally raises the question: which of these interpretations correctly describes physical reality?

Ultimately, all physical measurements reduce to position measurements, since experimental outcomes are recorded as approximate positions of macroscopic objects. Suppose, in principle, we perform an idealized measurement, determining the exact positions of all particles in a system (including those in the measuring apparatus) at a given time. Although such complete measurements are impractical, they provide a theoretical standard.  
In this case, if quantum mechanics and stochastic mechanics predict the same probability distribution for particle positions, they become experimentally indistinguishable.  
Indeed, as previously discussed, both theories yield the same probability density $\rho = |\psi|^2$ at a given time (cf.~Eq.~\eqref{map1}).

Nevertheless, their physical interpretations differ significantly.  
To illustrate this, consider the hydrogen atom in its ground state.  
In Coulomb units (where $m = e^2 = \hbar = 1$), the potential energy is $V = -1/r$, with $r = |x|$ the radial distance to the origin.  
The ground state wavefunction is
\begin{equation}
\psi(r) = \frac{1}{\sqrt{\pi}} e^{-r}.
\end{equation}
In quantum mechanics, this wavefunction provides a complete description of the system.  
In contrast, in stochastic mechanics, the electron follows a Markov process governed by the stochastic differential equation:
\begin{align}
dx(t) &= -\frac{x(t)}{|x(t)|} \, dt + dw(t) \\
&= b(x(t)) \, dt + dw(t),
\label{b}
\end{align}
where $w(t)$ is a Wiener process with diffusion coefficient $1/2$.  
Thus, the electron undergoes a highly irregular random motion, with a persistent drift toward the origin regardless of the direction of time—a behavior reminiscent of ordinary diffusion.

The question arises: how can such a classical stochastic model reproduce all observable predictions of quantum mechanics?  
The key is that, while in quantum mechanics the position operators at different times do not commute—and thus cannot simultaneously represent classical random variables—the trajectories in stochastic mechanics are random variables at each instant.

To clarify how no contradiction arises, consider a simpler example: a free particle with $m = \hbar = 1$, and an initial wavefunction
\begin{equation}
\psi_0(x) = N e^{-|x|^2/2a},
\end{equation}
where $N$ is a normalization constant.  
The time-evolved wavefunction is
\begin{equation}
\psi(x,t) = N e^{-\frac{|x|^2}{2(a + it)}} = N e^{-\frac{|x|^2 (a - it)}{a^2 + t^2}}.
\end{equation}
From the Nelson map [Eqs.~\eqref{map1}], one obtains the osmotic and current velocities:
\begin{align}
u(x,t) &= -\frac{a}{a^2 + t^2} x, \\
v(x,t) &= \frac{t}{a^2 + t^2} x,
\end{align}
and thus the drift velocity is
\begin{equation}
b(x,t) = u(x,t) + v(x,t) = \frac{t-a}{a^2+t^2} x.
\end{equation}
The particle thus follows the stochastic process:
\begin{equation}
dx(t) = \frac{t-a}{a^2+t^2} x \, dt + dw(t),
\end{equation}
where $w(t)$ is again a Wiener process with diffusion coefficient $1/2$.

Meanwhile, for a free particle in quantum mechanics, one has the relation:
\begin{equation}
x\left( \frac{t_1 + t_2}{2} \right) = \frac{x(t_1) + x(t_2)}{2},
\label{qm}
\end{equation}
valid for all times $t_1$ and $t_2$.  
However, in stochastic mechanics, such a deterministic relation between random variables $X(t)$ at different times does not hold.  
Thus, the mathematical structures of quantum mechanics and stochastic mechanics are fundamentally different.

Despite this, the theories remain experimentally indistinguishable.  
Any attempt to verify Eq.~\eqref{qm} operationally would involve measurements that, due to the uncertainty principle, introduce disturbances of the same order as the deviations predicted by the stochastic theory.  
Although Eq.~\eqref{qm} is formally valid for operators in quantum mechanics, it lacks operational meaning at the level of actual measurements.

Stochastic mechanics is, in this sense, conceptually simpler than quantum mechanics.  
In particular, it eliminates the paradoxes associated with wavefunction collapse and measurement.  
While the Schr\"{o}dinger evolution in quantum mechanics is deterministic and randomness enters only via measurement postulates (often involving the consciousness of the observer as a {\em deus ex machina}), stochastic mechanics is inherently indeterministic.  
Indeed, it has been shown~\cite{pav2} that wavefunction collapse can be derived from standard probability theory (via Bayes' theorem) and a stochastic variational principle, rather than postulated separately.  
After a measurement, the system follows a new Nelson stochastic process corresponding to the updated solution of the Schr\"{o}dinger equation.

Nelson himself recognized potential difficulties with stochastic mechanics, notably that Markov processes naturally live in configuration space, leading to concerns about nonlocality if particle paths are treated as physically real~\cite{nel2}.  
In particular, he discussed apparent contradictions between stochastic and quantum predictions for compound systems with entanglement.  
However, subsequent work~\cite{pet} showed that these issues can be resolved.  
By interpreting stochastic mechanics as a hidden variable theory for a finite number of spinless particles, and rigorously analyzing entangled systems, it was demonstrated that stochastic mechanics and quantum mechanics yield identical observable correlations at all times.

Thus, stochastic mechanics not only reproduces all quantum predictions, but also offers an alternative, realistic interpretation of microphysical processes.

\begin{table}[h]
\centering
\resizebox{\columnwidth}{!}{%
\begin{tabular}{|l|c|c|}
\hline
\textbf{Feature} & \textbf{Standard QM} & \textbf{Stochastic Mechanics} \\
\hline
Postulate wavefunction? & Yes & No \\
\hline
Particle trajectory? & Undefined & Defined via SDE \\
\hline
Source of randomness & Measurement only & Built into motion \\
\hline
Origin of Schr\"{o}dinger eq. & Fundamental axiom & Derived from Newton + Brownian motion \\
\hline
\end{tabular}%
}
\caption{Comparison between standard quantum mechanics and Nelson's stochastic mechanics.}
\label{tab:qm_vs_stoch}
\end{table}

In the following sections, we extend these ideas to field-theoretic systems. We demonstrate how the stochastic quantization framework can be systematically applied to photons and gravitons, thereby establishing a unified stochastic description for both quantum electrodynamics and linearized gravity.

\subsection {Relativistic Stochastic Mechanics}
The theory has also been extended to special relativistic physics by considering Markov processes in phase space, where well-defined stochastic evolutions are possible, as opposed to direct processes in Minkowski spacetime~\cite{dunk, garb, lind, pap1, yang, yor}. 
Another important development involves the use of analytic continuation techniques in Minkowski space.

A particularly interesting approach is the use of Poisson processes in spacetime instead of Wiener processes, combined with analytic continuation, to derive the Dirac equation~\cite{gav} as a quantum mechanical wave equation.  
In the following, we show that the Maxwell equations and the linearized Einstein equations, when recast in Schr\"{o}dinger-like form, can also be derived from an underlying Poisson process.

\section{Electrodynamics: Photon Wave Function}
\label{sec:photon}

To apply stochastic mechanics to electrodynamics, it is first necessary to formulate an appropriate quantum mechanical equation for the photon.  
We begin with Maxwell's equations in vacuum:
\begin{align}
\partial_t \mathbf{E}(\mathbf{r},t) &= c\, \nabla \times \mathbf{B}(\mathbf{r},t), \quad \nabla \cdot \mathbf{E}(\mathbf{r},t) = 0, \\
\partial_t \mathbf{B}(\mathbf{r},t) &= -c\, \nabla \times \mathbf{E}(\mathbf{r},t), \quad \nabla \cdot \mathbf{B}(\mathbf{r},t) = 0.
\end{align}
Defining the Riemann–Silberstein vectors:
\begin{equation}
\mathbf{F}_{\pm}(\mathbf{r},t) = \mathbf{E}(\mathbf{r},t) \pm i \mathbf{B}(\mathbf{r},t),
\label{RS}
\end{equation}
the Maxwell equations can be rewritten compactly as~\cite{biy}:
\begin{align}
i \partial_t \mathbf{F}_{\pm}(\mathbf{r},t) &= \pm c\, \nabla \times \mathbf{F}_{\pm}(\mathbf{r},t), \label{maxwell_RS} \\
\nabla \cdot \mathbf{F}_{\pm}(\mathbf{r},t) &= 0.
\end{align}

Using the identity between vector cross product and spin-1 matrix representations:
\begin{equation}
\mathbf{a} \times \mathbf{b} = -i (\mathbf{a} \cdot \mathbf{s}) \mathbf{b},
\end{equation}
where $(s_i)_{jk} = -i \epsilon_{ijk}$ are the spin-1 matrices generating the adjoint representation of $su(2)$, Eq.~\eqref{maxwell_RS} can be recast into the Schr\"{o}dinger-like form:
\begin{equation}
i\hbar \partial_t \mathbf{F}_{\pm}(\mathbf{r},t) = \mp i\hbar c\, \mathbf{s} \cdot \nabla\, \mathbf{F}_{\pm}(\mathbf{r},t).
\label{pw}
\end{equation}

This equation closely resembles the Weyl equation for massless neutrinos~\cite{weyl}:
\begin{equation}
i\hbar \partial_t \phi(\mathbf{r},t) = -i\hbar c\, \boldsymbol{\sigma} \cdot \nabla\, \phi(\mathbf{r},t),
\end{equation}
except that here the spin operators correspond to spin-1 representations rather than the spin-1/2 Pauli matrices.

In quantum mechanics, stationary solutions play a fundamental role.  
Separating variables as $\mathbf{F}_{\pm}(\mathbf{r},t) = \mathbf{F}_{\pm}(\mathbf{r}) e^{-i\omega t}$, Eq.~\eqref{pw} yields the eigenvalue equation:
\begin{equation}
c\, \mathbf{s} \cdot \hat{\mathbf{p}}\, \mathbf{F}_{\pm}(\mathbf{r}) = \hbar \omega\, \mathbf{F}_{\pm}(\mathbf{r}),
\label{eig}
\end{equation}
where $\hat{\mathbf{p}} = -i\hbar \nabla$ is the momentum operator.  
This equation describes a photon with positive helicity (for $\mathbf{F}_+$) or negative helicity (for $\mathbf{F}_-$).

The two helicity components $\mathbf{F}_\pm$ can be combined into a single six-component wavefunction:
\begin{equation}
\psi(\mathbf{r},t) = 
\begin{pmatrix}
\mathbf{F}_+(\mathbf{r},t) \\
\mathbf{F}_-(\mathbf{r},t)
\end{pmatrix}.
\label{six}
\end{equation}

%The physical photon wavefunction $\psi(\mathbf{r},t)$ is defined as the positive-frequency part of the electromagnetic field $\mathcal{F}(\mathbf{r},t)$:
%\[
%\psi(\mathbf{r},t) = \mathbf{F}_+(\mathbf{r},t).
%\]
%Since the negative-frequency part is the complex conjugate of the positive-frequency component, $\psi$ contains the full physical information.

The equation governing $\psi$ in free space is:

\begin{equation}
i\hbar \, \partial_t \psi(\mathbf{r},t) = c\, (\mathbf{\Sigma} \,\cdot \hat{\mathbf{p}}) \psi(\mathbf{r},t),
\label{sch1}
\end{equation}
where
\begin{equation}
\mathbf{\Sigma}_{i} =
\begin{pmatrix}
\mathbf{s}_{i}  & 0 \\
0 & -\mathbf{s}_{i} 
\end{pmatrix}.
\label{Sigma3_def}
\end{equation}

%\begin{equation}
%i\hbar \, \partial_t \psi(\mathbf{r},t) = c\, \Sigma_3 \, \hat{\mathbf{p}} \cdot \mathbf{s} \, \psi(\mathbf{r},t).
%\label{sch1}
%\end{equation}

%and $\mathbf{s}$ denotes the spin-1 vector matrices.

Detailed arguments establishing the normalizability and probability interpretation of $\psi$ can be found in~\cite{biy}.

Although the photon is fundamentally massless, we now introduce a small, 
nonzero mass-like term in the evolution equation for $\psi(\mathbf{r}, t)$. 
This does not alter the physical content, but serves as a mathematical tool 
that simplifies the stochastic formulation to follow. 
In particular, the mass term lifts the degeneracy between the helicity sectors 
that occurs when both $F_{+}$ and $F_{-}$ independently satisfy the same dispersion 
relation. Without such coupling, the six-component photon wavefunction contains 
a redundancy that obstructs a consistent probabilistic interpretation and 
dynamical framework. 

This construction is directly analogous to the role of the photon mass in 
Gupta--Bleuler quantization of electrodynamics, where a small mass is introduced 
to regularize the theory in a Lorentz-covariant form, with the physical content 
recovered by imposing a subsidiary condition and taking the massless limit. 
Here, the singular point $m_p = 0$ is excluded from the stochastic construction; 
all intermediate steps are carried out with $m_p \neq 0$. 
The mass term thus acts as a regulator enabling a well-defined stochastic dynamics, 
and the physical massless-photon predictions are recovered only at the end by 
taking the limit $m_p \to 0$.

With this motivation, we extend Eq.~\eqref{sch1} to the form:
\begin{equation}
i\hbar \partial_t \psi(\mathbf{r},t) = c (\boldsymbol{\Sigma} \cdot \hat{\mathbf{p}}) \psi(\mathbf{r},t) + m_{p} c^2 \beta \psi(\mathbf{r},t),
\label{sch2}
\end{equation}
where
\begin{equation}
\beta =
\begin{pmatrix}
0 & \mathbb{I}_3 \\
\mathbb{I}_3 & 0
\end{pmatrix}
\end{equation}
is a matrix that mixes the positive and negative helicity components, analogous to the Dirac $\beta$ matrix.

In the one-dimensional reduction of Eq.~\eqref{sch2}, where the wavefunction becomes a two-component field, the equation simplifies to:
\begin{equation}
i\hbar \frac{\partial \psi(x,t)}{\partial t} = -i c\hbar \Sigma_{1} \partial_x \psi(x,t) + m_{p} c^2 \beta \psi(x,t),
\label{1d}
\end{equation}
where $\Sigma_1$ corresponds to the spin-1 matrix in the $x$-direction.

\vspace{0.2cm}
\noindent

\subsection{Stochastic Quantization}

Following Ref.~\cite{gav}, consider a point particle with helicity~\cite{bar} propagating in empty space with velocity $v$ along the $x$-direction. The particle randomly flips both its direction (by $180^\circ$) and helicity at a rate determined by a parameter $m_{p}$. Let these reversals be Poisson-distributed, i.e., there is a fixed rate $a$ such that the probability of a flip in a time interval $dt$ is $a\,dt$. Let $P_+(x,t)$ and $P_-(x,t)$ denote the probability densities for the particle being at $x$ at time $t$, moving to the right and left, respectively.

A master equation over an infinitesimal time interval $\Delta t$ gives:
\begin{equation}
P_{\pm}(x,t+\Delta t) = P_{\pm}(x \mp \Delta x, t)(1 - a\Delta t) + P_{\mp}(x \pm \Delta x, t)a\Delta t.
\end{equation}
Taking the continuous-time limit leads to
\begin{equation}
\frac{\partial P_{\pm}}{\partial t} = -a(P_{\pm} - P_{\mp}) \mp v \frac{\partial P_{\pm}}{\partial x}, \qquad v = \left| \frac{\Delta x}{\Delta t} \right|,
\label{eq:X}
\end{equation}
and subsequently to the telegrapher's equation:
\begin{equation}
\frac{\partial^2 P_{\pm}}{\partial t^2} - v^2 \frac{\partial^2 P_{\pm}}{\partial x^2} = -2a \frac{\partial P_{\pm}}{\partial t}.
\label{eq:telegrapher}
\end{equation}

We now reinterpret the telegrapher’s equation in terms of a probability amplitude by defining
\begin{equation}
   \sqrt{\rho_{\pm}} = \ell_P^{3/2} P_{\pm},
   \label{hx}
\end{equation}
where $\ell_P$ is the Planck length. This rescaling ensures that
$\rho_{\pm}$ has dimensions of probability density and facilitates a comparison with relativistic quantum wave equations.

Multiplying Eqs.~(\ref{eq:X}) and~(\ref{eq:telegrapher}) by $\ell_P^{3/2}$, we obtain:
\begin{align}
\frac{\partial \sqrt{\rho_{\pm}}}{\partial t} &= - a (\sqrt{\rho_{\pm}} - \sqrt{\rho_{\mp}}) \mp v \frac{\partial \sqrt{\rho_{\pm}}}{\partial x}, \label{eq:rho1} \\
\frac{\partial^2 \sqrt{\rho_{\pm}}}{\partial t^2} - v^2 \frac{\partial^2 \sqrt{\rho_{\pm}}}{\partial x^2} &= - 2a \frac{\partial \sqrt{\rho_{\pm}}}{\partial t}. \label{eq:rho2}
\end{align}

Equation~(\ref{eq:rho1}) is structurally identical to the coupled first-order evolution equations for the helicity components of relativistic wavefields.  
To make this explicit, compare Eq.~(\ref{eq:rho1}) to the 1D Weyl/Dirac form
\[
\frac{\partial \phi_{\pm}}{\partial t} = \mp c\,\frac{\partial \phi_{\pm}}{\partial x} + \frac{i\,m_p c^2}{\hbar}(\phi_{\pm} - \phi_{\mp}) ,
\]
and identify
\begin{equation}
c \;\longleftrightarrow\; v, \qquad a \;\longleftrightarrow\; \frac{i\, m_{p} c^2}{\hbar}, \qquad \phi_{\pm} \;\longleftrightarrow\; \sqrt{\rho_{\pm}} .
\label{eq:rho-id}
\end{equation}
With this substitution, Eq.~(\ref{eq:rho1}) becomes exactly the relativistic helicity-component equation, showing that our stochastic amplitudes $\sqrt{\rho_{\pm}}$ obey the same dynamical form as $\phi_{\pm}$ in the quantum theory.

To extract the relativistic phase structure, we define  
\begin{equation}
\phi_{\pm}(x,t) := e^{-i m_{p} c^2 t / \hbar} \,\sqrt{\rho_{\pm}} .
\label{eq:phi-rho}
\end{equation}
The exponential factor is the rest-energy phase $e^{i S / \hbar}$ with $S = - m_{p} c^2 t$, just as in classical relativistic mechanics.  
This parallels Nelson's stochastic mechanics, where the modulus encodes stochasticity and the phase encodes deterministic action.

In the classical telegrapher’s equation, the term $-2a \,\partial_t \sqrt{\rho_{\pm}}$ is real-valued and causes exponential damping.  
Here, $a = i m_p c^2/\hbar$ is purely imaginary: analytic continuation changes the sign of the frequency term in Fourier space, converting decay $e^{-at}$ into oscillations $e^{-i\omega t}$. The resulting dynamics are dispersive rather than dissipative, producing sustained, phase-coherent wave motion.

\subsection*{Extension to three dimensions}
To generalize Eq.~(\ref{eq:X}) to 3D, we follow Ref.~\cite{gav}. Let $A$ be a linear generator of spatial translations (e.g., $A = -v\partial_x$ in 1D), and $N_a(t)$ a Poisson random variable with
\[
\text{Prob}[N_a(t+\epsilon) = N_a(t) + 1] = a\epsilon + \mathcal{O}(\epsilon^2), \qquad N_a(0) = 0.
\]
Conditional on the sign of $(-1)^{N_a(t)}$, the evolution operators are  
\begin{equation}
f_{\pm}(t) = \mathbb{E}\left[ \exp(s(t)A) \,\big|\, (-1)^{N_a(t)} = \pm 1 \right],
\end{equation}
where $f_{\pm}$ are now \emph{propagator operators} (scalars in 1D, $3\times 3$ matrices in the spin-1 case) acting on the helicity components. The corresponding master equation reads
\begin{equation}
\frac{\partial f_{\pm}}{\partial t} = -a(f_{\pm} - f_{\mp}) \pm A f_{\pm}.
\label{eq:gen}
\end{equation}

For a spin-1 field, we replace $A = -v \partial_x$ with the spin-1 translation generator $A = -c\,\mathbf{s} \cdot \nabla$, where $\mathbf{s}$ are the $3\times 3$ spin-1 matrices.  
The helicity components of the photon wavefunction are denoted $\mathbf{F}_\pm(\mathbf{r},t)$, which are the 3D analogues of the scalar $\phi_\pm$ in 1D.  
The six-component field
\[
\psi(\mathbf{r},t) = \begin{pmatrix} \mathbf{F}_+(\mathbf{r},t) \\ \mathbf{F}_-(\mathbf{r},t) \end{pmatrix}
\]
then evolves under $f_\pm$ via  
\[
\psi = e^{-i m_{p} c^2 t / \hbar} 
\begin{pmatrix} f_+\,\mathbf{F}_+ \\ f_-\,\mathbf{F}_- \end{pmatrix}.
\]
Substituting into Eq.~(\ref{eq:gen}) with $A = -c\, \mathbf{s} \cdot \nabla$ recovers Eq.~(\ref{sch2}), showing that the stochastic construction carries over seamlessly from massive spin-1/2 to massive spin-1 fields.  
The photon equation emerges in the limit $m_{p} \to 0$, $v \to c$.

Interestingly, in Eq.~(\ref{sch2}), Planck’s constant \(\hbar\) cancels out entirely, similarly to the Weyl equation for massless spin-\(\tfrac{1}{2}\) particles. The resulting equation,
\begin{equation}
i \partial_t \psi(\mathbf{r}, t) = -i c\, (\boldsymbol{\Sigma} \cdot \nabla) \psi(\mathbf{r}, t),
\label{eq:classical}
\end{equation}
is structurally equivalent to the classical Maxwell equations expressed in terms of the Riemann–Silberstein vector in an inhomogeneous medium. Unlike the stationary eigenvalue form Eq.~(\ref{eig}), however, Eq.~(\ref{eq:classical}) does not represent a quantized energy eigenvalue equation. In contrast, Eq.~(\ref{sch1}) describes a genuinely quantum-mechanical wave equation for photons with quantized energy \(\hbar \omega\). The classical dynamics of Eq.~(\ref{eq:classical}) is embedded within the quantum theory as a special limiting case, without requiring \(\hbar = 0\).

This stochastic derivation of the photon wavefunction recovers a structure well-known in quantum optics, providing a natural bridge to second-quantized formulations\footnote{In second-quantized form, the electromagnetic field operator can be written as
\[
\hat{\psi}(\mathbf{r}, t) = \sum_n \left[ \Psi_n(\mathbf{r}, t) \hat{c}_n + \rho_1 \Psi_n^*(\mathbf{r}, t) \hat{c}_n^\dagger \right],
\]
where \(\Psi_n(\mathbf{r}, t)\) are photon mode functions obtained from the first-quantized equation and \(\rho_1\) is a Pauli matrix acting on the helicity structure~\cite{biy}. This construction underlies the optical equivalence theorem~\cite{sud} and highlights the compatibility of our stochastic approach with standard quantum optical field operators.}.
This also helps explain why quantum-like features such as mode entanglement and coherence can arise in classical wave systems despite the absence of particle localization.

We emphasize, however, that our notion of “stochastic electrodynamics/optics” differs fundamentally from that of Marshall~\cite{mar}, Boyer~\cite{boy}, and others~\cite{pen}, who augment the classical Maxwell field with ad hoc zero-point radiation to mimic quantum fluctuations. In contrast, our approach derives the quantum mechanical wave equation itself from a stochastic process governing a classical spin-1 particle, without requiring externally imposed vacuum noise.\\

\section{Weak Gravity: Graviton Wave Function}

In this section, we formulate the stochastic quantization of linearized gravity. The pure Einstein–Hilbert action in $3+1$ dimensions is given by
\begin{equation}
S_{EH} = - \int d^4x\, \mathcal{L}_E,
\end{equation}
where the Einstein Lagrangian $\mathcal{L}_E$ is defined as
\begin{equation}
\mathcal{L}_E = \sqrt{-g} R = \sqrt{-g}\, g^{\mu\nu} R_{\mu\nu},
\label{eq:full_Einstein_Lagrangian}
\end{equation}
with \( g = \det(g_{\mu\nu}) \) the determinant of the metric, and \( R_{\mu\nu} \) the Ricci tensor.

To proceed, we begin by deriving the classical action for a weak gravitational field.  In the weak-field regime, the metric can be written as
\begin{equation}
g_{\mu\nu} = \eta_{\mu\nu} + h_{\mu\nu},
\end{equation}
where \( \eta_{\mu\nu} = \text{diag}(1, -1, -1, -1) \) is the Minkowski metric in Cartesian coordinates, and \( |h_{\mu\nu}| \ll 1 \) represents small perturbations.

The linearized form of the Einstein–Hilbert Lagrangian, expanded around flat Minkowski space, reads
\begin{equation}
\mathcal{L}_{\text{EH}} = \frac{1}{2} h^{\mu\nu} \left( R_{\mu\nu}^{(L)} - \frac{1}{2} \eta_{\mu\nu} R^{(L)} \right),
\label{eq:linearized_EH_Lagrangian}
\end{equation}
where \( R_{\mu\nu}^{(L)} \) is the linearized Ricci tensor:
\begin{equation}
R_{\mu\nu}^{(L)} = \frac{1}{2} \left( -\Box h_{\mu\nu} + \partial_\mu \partial^\alpha h_{\alpha\nu} + \partial_\nu \partial^\alpha h_{\alpha\mu} - \partial_\mu \partial_\nu h \right),
\label{eq:linearized_Ricci}
\end{equation}
with \( h = h^\alpha_{\ \alpha} \) the trace of the perturbation. The linearized Ricci scalar is
\begin{equation}
R^{(L)} = \eta^{\mu\nu} R_{\mu\nu}^{(L)}.
\end{equation}

Varying the action corresponding to Eq.~\eqref{eq:linearized_EH_Lagrangian} with respect to \( h_{\mu\nu} \) yields the field equations:
\begin{equation}
-\Box h_{\mu\nu} + \partial_\mu \partial^\alpha h_{\alpha\nu} + \partial_\nu \partial^\alpha h_{\alpha\mu} - \partial_\mu \partial_\nu h = 0.
\label{gd}
\end{equation}

Define the trace-reversed perturbation as
\begin{equation}
\bar{h}_{\mu\nu} = h_{\mu\nu} - \frac{1}{2}\eta_{\mu\nu}h,
\end{equation}
where \( h = \eta^{\mu\nu}h_{\mu\nu} \). Imposing the harmonic (de Donder) gauge condition,
\begin{equation}
\partial^\nu \bar{h}_{\mu\nu} = 0,
\end{equation}
the field equations simplify to
\begin{equation}
\Box \bar{h}_{\mu\nu} = 0.
\end{equation}

However, the harmonic gauge does not uniquely determine \( h_{\mu\nu} \). Under the gauge transformation
\begin{equation}
h_{\mu\nu} \rightarrow h_{\mu\nu} + \partial_\mu \zeta_\nu + \partial_\nu \zeta_\mu,
\label{n}
\end{equation}
with \( \Box \zeta_\mu = 0 \), the harmonic gauge condition remains invariant. Therefore, to uniquely fix \( h_{\mu\nu} \), four additional conditions must be imposed.

Introducing a unit timelike vector \( u^\mu \) associated with an observer detecting the gravitational wave, we define the \textbf{transverse-traceless (TT)} gauge. In this gauge, the metric perturbation satisfies
\begin{equation}
h_{0\mu} = 0, \quad \partial_i h_{ij} = 0, \quad h^\mu_{\ \mu} = 0.
\label{TT}
\end{equation}
The transversality condition \( \partial_i h_{ij} = 0 \) is observer-dependent, while the tracelessness condition \( h^\mu_{\ \mu} = 0 \) implies \( \bar{h} = 0 \), so that \( \bar{h}_{\mu\nu} = h_{\mu\nu} \) in this gauge. The wave equation in TT gauge then becomes
\begin{equation}
\Box h_{ij}(t, \vec{x}) = 0, \quad i,j = 1,2,3.
\end{equation}

Since \( h_{ij} \) is a symmetric, traceless rank-2 tensor in three dimensions, it is convenient to decompose it into spherical tensor components \( h^{(2)}_q \), with \( q = -2, -1, 0, +1, +2 \), corresponding to the helicity states of a spin-2 field~\cite{pn2}. These modes represent different angular momentum projections along the direction of wave propagation.

In the TT gauge, many components vanish due to transversality and tracelessness. In particular, for a gravitational wave propagating along the \( z \)-axis, only the Cartesian components \( h_{11} \) and \( h_{12} \) are nonzero. These correspond to the two physical polarizations of the gravitational wave (see Appendix).

The spherical and Cartesian components \cite{pn2,pn3,pn1} are related as:
\begin{align}
h_{11} &= \frac{1}{2} \left( h^{(2)}_{+2} + h^{(2)}_{-2} \right), \\
h_{22} &= -\frac{1}{2} \left( h^{(2)}_{+2} + h^{(2)}_{-2} \right), \\
h_{12} &= \frac{i}{2} \left( h^{(2)}_{+2} - h^{(2)}_{-2} \right), \\
h_{33} &= h^{(2)}_0 = - (h_{11} + h_{22}) = 0, \\
h_{13} &= - \frac{1}{\sqrt{2}} \left( h^{(2)}_{+1} - h^{(2)}_{-1} \right) = 0, \\
h_{23} &= \frac{i}{\sqrt{2}} \left( h^{(2)}_{+1} + h^{(2)}_{-1} \right) = 0.
\end{align}

The vanishing of the \( h^{(2)}_0 \) and \( h^{(2)}_{\pm1} \) modes for a wave propagating along the  z-axis in TT gauge confirms that only the \( q = \pm 2 \) helicities are physical, corresponding to the two dynamical degrees of freedom of the grvitational modes.

The wave equation for the spherical tensor components in free space is
\begin{equation}
\Box\, h^{(2)}_q(\vec{x}, t) = 0,
\label{eq:gw_wave}
\end{equation}
where \( q = 0, \pm 1, \pm 2 \). In TT gauge, only the helicity modes \( q = \pm 2 \) survive, and \( h^{(2)}_q \) depends only on \( z \), consistent with the wave propagating along the \( z \)-axis.

To isolate these physical degrees of freedom, and in analogy with the photon case, we define a two-component helicity spinor:
\begin{equation}
\psi^{(2)}_{g}(z, t) = \begin{pmatrix}
h^{(2)}_{+2}(z, t) \\
h^{(2)}_{-2}(z, t)
\end{pmatrix},
\end{equation}

which spans the physical subspace of the full five-component spin-2 field \( h^{(2)}(\vec{x}, t) = \sum_{q=-2}^{2} h_q^{(2)}(\vec{x}, t) |2, q\rangle \). This spinor captures the two propagating polarization modes of the gravitational wave in a compact form.

By analogy with the photon, one can formulate a first-order Schr\"{o}dinger-like evolution equation \`a la Dirac for the graviton helicity spinor in the transverse-traceless (TT) gauge:
\begin{equation}
i\hbar \frac{\partial \psi^{(2)}_{g}(z, t)}{\partial t}
= -i\hbar c\, (\boldsymbol{\sigma} \cdot \nabla)\, \psi^{(2)}_{g}(z, t),
\label{eq:graviton_spinor}
\end{equation}
where \( \boldsymbol{\sigma} = (\sigma_x, \sigma_y, \sigma_z) \) are the Pauli matrices. The Pauli matrices here do not represent spin per se but rather encode the dynamics within a two-dimensional subspace. While a massless spin-2 field possesses five helicity modes, the TT gauge restricts attention to the physical helicities \( q = \pm 2 \). Within this reduced subspace, the Pauli matrices act as effective generators of spatial rotations, analogous to spin-2 operators confined to the physical sector. This formulation offers a natural basis for implementing stochastic quantization in linearized gravity.

\section{Stochastic Quantization of Linearized Gravity: Graviton Case}

After imposing the transverse-traceless (TT) gauge --- where the conditions \( h_{0\mu} = 0 \), \( \partial^i h_{ij} = 0 \), and \( h^i_{\ i} = 0 \) hold --- the set of allowable gauge transformations is restricted to those \( \zeta^\mu \) that preserve these conditions. Specifically, we require:
\begin{align}
\delta h_{0\mu} &= \partial_0 \zeta_\mu + \partial_\mu \zeta_0 = 0, \\
\delta (\partial^i h_{ij}) &= \Box \zeta_j + \partial_j (\partial^i \zeta_i) = 0, \\
\delta (h^i_{\ i}) &= 2 \partial^i \zeta_i = 0.
\end{align}
These imply the residual gauge transformations must satisfy:
\begin{align}
\Box \zeta^i &= 0, \\
\partial^i \zeta_i &= 0, \\
\partial_0 \zeta_i &= -\partial_i \zeta_0, \\
\partial_0 \zeta_0 &= 0.
\end{align}
Such transformations leave the TT gauge conditions invariant, and consequently, the transverse-traceless components \( h_{+} \) and \( h_{\times} \) remain unchanged under them. These components thus represent true, gauge-invariant physical degrees of freedom.

To formulate a consistent stochastic evolution equation for the graviton field in a first-order Schr\"{o}dinger-like form, we introduce an auxiliary mass term in the helicity-2 sector:
\begin{equation}
   i\hbar \frac{\partial \psi_g(t,z)}{\partial t} = -i\hbar c\,\sigma_{3} \frac{\partial \psi_g}{\partial z} + m_g c^2\, \sigma_1 \psi_g, 
   \label{hg}
\end{equation}
where \( \sigma_{1,3} \) are Pauli matrices acting on the helicity components \( \pm 2 \). This mass term facilitates the stochastic dynamics (via a Poisson process analogy) by coupling the helicity sectors, but it does not introduce new physical degrees of freedom or affect gauge-invariant observables in the TT sector. This is not derived from GR, but postulated as an effective first-order system whose solutions reduce to the correct second-order wave equation in the massless limit.

In our formulation, the auxiliary graviton mass $m_g$ serves purely as a 
mathematical regulator that couples the helicity sectors and enables a 
consistent stochastic construction. The singular point $m_g = 0$ is excluded; 
all derivations are carried out with $m_g \neq 0$. The physical massless 
spin-2 theory in the TT gauge is recovered only at the end, by taking 
the limit $m_g \to 0$.

%This construction is closely analogous to the role of the photon mass in Gupta--Bleuler quantization of electrodynamics. There, a small mass term is introduced to regularize the theory in a Lorentz-covariant form, and the physical content is recovered by imposing a subsidiary condition and taking the massless limit. Similarly, in our case, we construct the stochastic graviton dynamics with the auxiliary mass and then take the limit \( m_g \to 0 \) to restore consistency with massless spin-2 field theory in the TT gauge.

\subsection{Helicity-2 Dynamics and Poisson Process}

In the transverse-traceless (TT) gauge, the linearized Einstein field equations reduce to a wave equation for the symmetric, traceless, and transverse metric perturbation \( h_{ij} \). These components can be expressed in terms of spin-2 spherical tensor components \( h_q^{(2)} \) with \( q = -2, -1, 0, +1, +2 \), where only the \( q = \pm 2 \) modes correspond to physical degrees of freedom.

To model these components stochastically, consider a spin-2 particle (graviton) moving along the \( z \)-axis at speed \( c \), with helicity states \( \pm 2 \). Let \( P^{g}_\pm(z, t) \) denote the probability densities for right/left-moving helicity-2 states. Assume that the graviton undergoes random helicity and direction flips governed by a Poisson process with rate \( a \). Then:
\begin{equation}
    \frac{\partial P^{g}_\pm}{\partial t} = -a(P^{g}_\pm - P^{g}_\mp) \mp c \frac{\partial P^{g}_\pm}{\partial z},
    \label{eq:masterg}
\end{equation}
leading to the telegrapher’s equation:
\begin{equation}
    \frac{\partial^2 P^{g}_\pm}{\partial t^2} - c^2 \frac{\partial^2 P^{g}_\pm}{\partial z^2} = -2a \frac{\partial P^{g}_\pm}{\partial t}.
    \label{eq:teleg}
\end{equation}

Now define the Planck-rescaled amplitude:
\begin{equation}
\sqrt{\rho^{g}_\pm} = \ell_P^{3/2} P^{g}_\pm,
\end{equation}
so that Eqs.~(\ref{eq:masterg})–(\ref{eq:teleg}) become:
\begin{align}
\frac{\partial \sqrt{\rho^{g}_\pm}}{\partial t} &= -a(\sqrt{\rho^{g}_\pm} - \sqrt{\rho^{g}_\mp}) \mp c \frac{\partial \sqrt{\rho^{g}_\pm}}{\partial z}, \label{eq:amp1} \\
\frac{\partial^2 \sqrt{\rho^{g}_\pm}}{\partial t^2} - c^2 \frac{\partial^2 \sqrt{\rho^{g}_\pm}}{\partial z^2} &= -2a \frac{\partial \sqrt{\rho^{g}_\pm}}{\partial t}. \label{eq:amp2}
\end{align}

As before, define a two-component helicity wavefunction via
\begin{equation}
\psi^{(2)}_g(z, t) = e^{-i m_g c^2 t/\hbar} \begin{pmatrix} \sqrt{\rho^g_+} \\ \sqrt{\rho^g_-} \end{pmatrix},
\end{equation}
which satisfies the Schr\"{o}dinger-type equation~(\ref{hg}) with the identification:
\begin{equation}
a = \frac{i m_g c^2}{\hbar}, \qquad u^g_{\pm 2} \equiv \sqrt{\rho^{g}_\pm}.
\end{equation}

This form again parallels the photon case: the stochastic dynamics of \( \sqrt{\rho^g_\pm} \) obeys a first-order coupled system that, in the \( m_g \to 0 \) limit, reduces to:
\begin{equation}
\square u^g_{\pm 2}(t, z) = 0,
\end{equation}
which is the massless spin-2 wave equation in vacuum.

\vspace{0.5em}
\noindent

As in the electromagnetic case, the key conceptual feature is that on analytic continuation to the complex plane, the flip rate \( a = \frac{im_g c^2}{\hbar} \) is \emph{purely imaginary}, and the classical dissipative process is converted into a \emph{dispersive, coherent helicity-mixing process}. Thus, quantum wave propagation for spin-2 emerges from an underlying stochastic dissipative dynamics on analytic continuation.

\vspace{0.5em}
\noindent
This construction further supports the broader idea that relativistic quantum equations—whether for spin-1 or spin-2 fields—can emerge from analytically continued stochastic dynamics, where helicity transitions mimic the internal structure of quantum fields.

In this framework, the quantum field nature of the graviton is encoded not in field operators but in the statistical amplitudes \( \sqrt{\rho^{g}_\pm} \), which satisfy a Klein–Gordon-type equation in the massless limit. The entire wave equation for linearized gravity is thus recovered as the statistical limit of a first-order, Poisson-driven process.

\section*{VI. Conclusion}

In this work, we have proposed an alternative formulation of quantum dynamics for massless gauge fields—specifically, the photon and the graviton—within the framework of stochastic mechanics, taking inspiration from Nelson’s approach to nonrelativistic quantum theory. Rather than applying canonical quantization, we suggest that quantum behavior may emerge from an underlying classical stochastic process. While Nelson’s model relies on Brownian motion and is well-suited to describing massive particles, our approach is built on a Poisson-type stochastic process that naturally accommodates massless excitations and their discrete helicity structure.

By applying our stochastic framework to electromagnetic and linearized gravitational fields, we derive Dirac-like evolution equations for helicity-resolved vector and tensor wavefunction-like quantities. Although these objects resemble conventional quantum wavefunctions, their probabilistic interpretation does not rely on spatial localization. 

Although our framework involves a small but nonzero mass—making the Newton–Wigner localization procedure formally applicable—the probabilistic interpretation of the wavefunction does not rely on sharply localized position eigenstates. Instead, particle position appears as a stochastic variable within an underlying Poisson process, and the probabilistic content arises dynamically from this process. Thus, our framework sidesteps the localization issues that plague conventional massless wavefunctions, while retaining a well-defined stochastic interpretation grounded in field amplitudes.

This leads to a fundamentally different perspective: the quantum behavior of ``gauge fields" is not rooted in postulated indeterminism--as in the Born rule in standard quantum mechanics--or particle localization, but in an underlying classical stochastic process with intrinsic probabilistic content.

%\textbf{Applying this framework to the electromagnetic and linearized gravitational fields, we derive Schr\"{o}dinger-type evolution equations for helicity-resolved vector and tensor wavefunction-like quantities. Although these objects formally resemble standard quantum wavefunctions, they do not support a probability interpretation based on spatial localization, since photons and gravitons lack well-defined position operators. In particular, the Newton–Wigner construction, which defines localized states for massive particles, fails for massless particles with nonzero helicity due to the structure of the Poincaré little group.}

%\textbf{In contrast to conventional quantum theory, where probability is introduced axiomatically via the Born rule, our framework allows probabilistic content to emerge dynamically from the underlying stochastic evolution. The squared modulus of the wavefunction encodes statistical weights over field configurations, shaped by Poissonian helicity-switching processes. Even though the helicity-flip rate vanishes in the physical (massless) limit, the probabilistic structure of the wavefunction persists, having been imprinted by the stochastic process before the limit is taken.}

%\textbf{Thus, probability in our construction is not linked to particle position, but to the dynamical evolution of field amplitudes. This leads to a picture in which the quantum behavior of gauge fields arises not from postulated indeterminism, but from classical stochastic dynamics.}

Our approach begins with the classical field equations of electrodynamics and linearized gravity and recasts them in Schr\"{o}dinger-like form. We then show that this form can be derived from a Poisson-type stochastic process, offering a natural probabilistic interpretation of classical field amplitudes. Crucially, this is not a quantization procedure in the conventional sense: rather than treating fields as operators in a Hilbert space, we interpret them as wavefunction-like entities whose statistical content arises from an underlying stochastic process. This framework offers a novel bridge between classical field theory and quantum-like dynamics, revealing how a probabilistic structure can emerge without invoking canonical quantization or the Born rule. In doing so, our work extends the scope of stochastic mechanics to encompass massless spin-1 and spin-2 fields, providing a new foundation for interpreting classical gauge fields as carriers of quantum-like statistical behavior.

To connect this stochastic dynamics with classical wave propagation, we introduced a small mass parameter that couples the helicity sectors and regularizes the first-order evolution equations. These equations lead to second-order telegraph-type equations, featuring complex damping terms. Upon analytic continuation, these damping terms acquire imaginary parts and transform classical dissipation into quantum-like dispersion. In the massless limit, the stochastic equations reduce to standard relativistic wave equations, revealing classical field equations as limiting cases of a deeper stochastic dynamics.

Importantly, while we take the limits \( m_p \to 0 \) and \( m_g \to 0 \) to recover the physical wave equations for photons and gravitons, we explicitly exclude the singular points \( m = 0 \) from the stochastic construction. This is because the underlying helicity-reversal mechanism, essential for generating wave-like behavior, vanishes identically when \( m = 0 \) is imposed from the outset. Nevertheless, the quantum mechanical structure of the wavefunction persists and remains well-defined in the limiting process. Physical observables remain smooth and finite, and the use of small but nonzero mass parameters is entirely consistent with current experimental constraints, so long as the values lie below observational bounds.

Notably, the telegraph structures derived for photons and gravitons exhibit a striking formal similarity, despite differences in gauge structure and tensorial rank. This suggests a universal underlying stochastic mechanism governing the propagation of massless gauge fields. The contrast between spin-1 and spin-2 dynamics is thereby encoded entirely in the symmetry content of the wavefunctions, not in the form of the stochastic evolution itself.

From a broader perspective, this work contributes to the growing line of inquiry in which quantum mechanics is reinterpreted as an emergent theory arising from deeper classical stochastic processes. The introduction of an effective constant analogous to Planck’s constant in our formulation allows a smooth transition from classical probabilistic dynamics to quantum-like field behavior. This provides a novel conceptual bridge between classical stochastic models and the structure of quantum field theory.

\begin{table*}[t]
\caption{Stochastic emergence of quantum dynamics: Photon vs Graviton}
\centering
\begin{ruledtabular}
\begin{tabular}{lcc}
\textbf{Aspect} & \textbf{Photon} & \textbf{Graviton} \\
\hline
Field content & Gauge-invariant electric and magnetic fields & Metric perturbation $h_{\mu\nu}$ in linearized gravity \\
Gauge treatment & Riemann–Silberstein vector $\vec{F}_{\pm} = \vec{E} \pm i \vec{B}$ & TT gauge isolates radiative degrees of freedom \\
Helicity basis & Helicity $\pm 1$ components of $\vec{F}_\pm$ & Helicity $\pm 2$ components of $h_{q}^{(2)}$ \\
Wavefunction & $\vec{F}_\pm$: complex vector encoding helicity states & $h^{(2)}_{\pm 2}$: graviton wavefunction in TT gauge \\
Stochastic dynamics & Poisson-type helicity-flip process & Same process applied to tensorial wavefunctions \\
Evolution equation & Schr\"{o}dinger-like equation for $\vec{F}_\pm$ & Schr\"{o}dinger-like equation for $h^{(2)}_{\pm 2}$ \\
Telegraph-type structure & Yes; reduces to $\Box \vec{F}_\pm = 0$ in massless limit & Yes; reduces to $\Box h^{(2)}_{\pm 2} = 0$ in massless limit \\
Physical interpretation & Emergent photon dynamics from stochasticity & Emergent graviton dynamics from stochasticity \\
Statistical picture & Probabilistic view via helicity flips & Analogous stochastic interpretation for gravitons \\
\end{tabular}
\end{ruledtabular}
\label{tab:stochastic_photon_graviton}
\end{table*}

In summary, the wave equations governing the propagation of massless ``gauge fields"—whether in electrodynamics or linearized gravity—emerge naturally from an underlying helicity-reversing stochastic process described by a Telegrapher-type equation, following analytic continuation.

%In summary, the wave equations that govern the propagation of massless gauge fields—whether in electrodynamics or linearized gravity—can be understood as the emergent limit of an underlying stochastic telegraph process. This framework provides a robust probabilistic interpretation of field dynamics, and offers a new lens through which to view the foundations of quantum field theory, the role of stochasticity, and the quantum–classical transition.

\section{Future Prospects}

Several key avenues for future research follow naturally from the stochastic helicity-reversal framework developed here:

\begin{itemize}
    \item Generalizing this stochastic helicity-reversal framework to electrodynamics in dielectric media via Pal’s covariant approach~\cite{Pal:2021}, enabling study of helicity-dependent effects such as the photon spin Hall effect in curved spacetime.

    \item Extending the stochastic formalism to the full Standard Model \cite{Nandi:prep}, where particle masses arise without spontaneous symmetry breaking and a Planck-scale cutoff ensures ultraviolet finiteness, leading to a realistic reformulation of quantum field theory.

    \item Exploring whether a stochastic process with fluctuating metrics \cite{Breuer:2009} can yield a nonlinear Einstein-like equation, offering a statistical emergence of spacetime geometry connected to loop quantum gravity frameworks \cite{Rovelli:2004tv,Perez:2012wv}.
\end{itemize}

Further theoretical and phenomenological investigations are underway and will be detailed in future publications.

\section{Acknowledgements}
PN acknowledges support from the Rector’s Postdoctoral Fellowship Program (RPFP) at Stellenbosch University. He is also grateful to Prof. Francesco Petruccione and Dr. Anslyn John for their insightful discussions and valuable feedback. PG thanks the Director and staff of IIT Mandi for their warm support and hospitality during his visit, when this work was completed.

\section{Appendix: Polarization Structure of Linearized Gravity}

We analyze the polarization structure of linearized gravitational waves using the plane-wave method, following the standard approach for gauge theories. The gauge parameters $\zeta_\mu(x)$ are assumed to be small and arbitrary. 

We adopt the ansatz for the metric perturbation:
\begin{equation}
h_{\mu\nu}(x) = \epsilon_{\mu\nu}(k)\, e^{ik \cdot x} + \text{c.c.},
\label{eq:ansatz_h}
\end{equation}
where $\epsilon_{\mu\nu}(k)$ is a symmetric polarization tensor. Correspondingly, the gauge parameter is chosen as
\begin{equation}
\zeta_\mu(x) = \zeta_\mu(k)\, e^{ik \cdot x} + \text{c.c.},
\label{eq:ansatz_zeta}
\end{equation}
so that the gauge transformation takes the form
\begin{equation}
\epsilon_{\mu\nu}(k) \rightarrow \epsilon'_{\mu\nu}(k) = \epsilon_{\mu\nu}(k) + k_\mu \zeta_\nu(k) + k_\nu \zeta_\mu(k).
\label{eq:gauge_eps}
\end{equation}

Substituting Eq.~\eqref{eq:ansatz_h} into the linearized Einstein field equations yields the momentum-space condition:
\begin{align}
k^2 \epsilon_{\mu\nu}
- k_\mu k^\alpha \epsilon_{\alpha\nu}
- k_\nu k^\alpha \epsilon_{\alpha\mu}
+ k_\mu k_\nu \epsilon
+ \eta_{\mu\nu} \left(-k^2 \epsilon + k^\alpha k^\beta \epsilon_{\alpha\beta} \right) = 0,
\label{eq:eom_momentum}
\end{align}
where $\epsilon \equiv \eta^{\mu\nu} \epsilon_{\mu\nu}$ is the trace of the polarization tensor.

We consider two cases separately:

\subsection*{1. Massive Case \texorpdfstring{(\(k^2 \neq 0\))}{(k² ≠ 0)}}

Contracting Eq.~\eqref{eq:eom_momentum} with $\eta^{\mu\nu}$ yields:
\begin{equation}
k^2 \epsilon - k^\alpha k^\beta \epsilon_{\alpha\beta} = 0.
\label{eq:trace_massive}
\end{equation}
A general solution takes the form
\begin{equation}
\epsilon_{\mu\nu}(k) = k_\mu f_\nu(k) + k_\nu f_\mu(k),
\label{eq:general_solution_massive}
\end{equation}
where $f_\mu(k)$ is arbitrary. This solution is pure gauge, as it can be eliminated by the choice $\zeta_\mu(k) = f_\mu(k)$. Therefore, as in Maxwell theory, all massive excitations are gauge artifacts and do not correspond to physical degrees of freedom in linearized gravity.

\subsection*{2. Massless Case \texorpdfstring{(\(k^2 = 0\))}{(k² = 0)}}

For massless modes, Eq.~\eqref{eq:eom_momentum} simplifies to:
\begin{equation}
- k_\mu k^\alpha \epsilon_{\alpha\nu}
- k_\nu k^\alpha \epsilon_{\alpha\mu}
+ k_\mu k_\nu \epsilon
+ \eta_{\mu\nu} k^\alpha k^\beta \epsilon_{\alpha\beta} = 0.
\label{eq:eom_massless}
\end{equation}

To analyze this, we choose the standard frame where the null momentum vector takes the form:
\begin{equation}
k^\mu = (\omega, 0, 0, \omega)^{T}.
\end{equation}
In this frame, Eq.~\eqref{eq:eom_massless} becomes:
\begin{multline}
- \omega \left[ 
k_\mu (\epsilon_{0\nu} - \epsilon_{3\nu}) 
+ k_\nu (\epsilon_{0\mu} - \epsilon_{3\mu}) 
\right]
+ k_\mu k_\nu \epsilon \\
+ \omega^2 \eta_{\mu\nu} (\epsilon_{00} + \epsilon_{33} - 2\epsilon_{03}) = 0.
\label{eq:eom_massless_explicit}
\end{multline}

Using the symmetry $\epsilon_{\mu\nu} = \epsilon_{\nu\mu}$, individual components of Eq.~\eqref{eq:eom_massless_explicit} yield the following constraints:
\begin{alignat}{2}
\epsilon_{11} &= -\epsilon_{22}, \qquad &\text{from } \mu = \nu = 0, \\
\epsilon_{01} &= -\epsilon_{31}, \qquad &\text{from } \mu = 1, \nu = 0, \\
\epsilon_{02} &= -\epsilon_{32}, \qquad &\text{from } \mu = 2, \nu = 0, \\
\epsilon_{03} &= -\epsilon_{33}, \qquad &\text{from } \mu = 3, \nu = 0, \\
\epsilon_{00} &= -\epsilon_{03}, \qquad &\text{from } \mu = \nu = 1.
\end{alignat}

From these conditions, the polarization tensor reduces to the form:
\begin{equation}
\{ \epsilon_{\mu\nu} \} =
\begin{pmatrix}
\epsilon_{00} & \epsilon_{01} & \epsilon_{02} & \epsilon_{00} \\
\epsilon_{01} & \epsilon_{11} & \epsilon_{12} & \epsilon_{01} \\
\epsilon_{02} & \epsilon_{12} & -\epsilon_{11} & \epsilon_{02} \\
\epsilon_{00} & \epsilon_{01} & \epsilon_{02} & \epsilon_{00}
\end{pmatrix}.
\label{eq:epsilon_reduced}
\end{equation}

Applying a gauge transformation with:
\begin{equation}
\zeta_0 = \zeta_3 = -\frac{\epsilon_{00}}{\omega}, \quad
\zeta_1 = -\frac{\epsilon_{01}}{\omega}, \quad
\zeta_2 = -\frac{\epsilon_{02}}{\omega},
\end{equation}
we eliminate the unphysical components and obtain the maximally reduced form:

\begin{equation}
\epsilon_{\mu\nu} =
\begin{pmatrix}
0 & 0 & 0 & 0 \\
0 & \epsilon_{+} & \epsilon_{\times} & 0 \\
0 & \epsilon_{\times} & -\epsilon_{+} & 0 \\
0 & 0 & 0 & 0
\end{pmatrix}
\quad \Rightarrow
h_{\mu\nu}(x) \sim 
\begin{pmatrix}
0 & 0 & 0 & 0 \\
0 & h_{11} & h_{12} & 0 \\
0 & h_{21} & -h_{11} & 0 \\
0 & 0 & 0 & 0
\end{pmatrix},
\label{eq:polarization_final}
\end{equation}
which is gauge-equivalent to Eq.~\eqref{eq:epsilon_reduced}.

In this transverse-traceless (TT) gauge, only the two physical polarization modes remain: $\epsilon_+$ and $\epsilon_\times$, corresponding to the two helicity states of a massless spin-2 field. The TT gauge satisfies:
\begin{align}
k^\mu \epsilon_{\mu\nu}(k) &= 0 \quad \text{(transversality)}, \label{eq:TT_transverse} \\
\epsilon^\mu_{\ \mu}(k) &= 0 \quad \text{(tracelessness)}, \label{eq:TT_traceless}
\end{align}
making it the standard and physically transparent choice for describing gravitational radiation.


\begin{thebibliography}{99}

\bibitem{Ashtekar198}
A.~Ashtekar, Phys.\ Rev.\ Lett.\ \textbf{57}, 2244 (1986).

\bibitem{1}
D.~G.~Boulware and S.~Deser, Ann.\ Phys.\ (N.Y.) \textbf{89}, 193 (1975).

\bibitem{2}
P.~Nandi and B.~R.~Majhi, Phys.\ Lett.\ B \textbf{857}, 138988 (2024), arXiv:2403.11253 [hep-th].
\bibitem{Sugiyama2024}
Y.~Sugiyama, A.~Matsumura, and K.~Yamamoto,
\emph{Phys. Rev. D} \textbf{110}, 045016 (2024).

\bibitem{2+}
B.~P.~Abbott \textit{et al.} (LIGO Scientific Collaboration and Virgo Collaboration), Phys.\ Rev.\ Lett.\ \textbf{116}, 061102 (2016).

\bibitem{3}
A.~J.~Leggett, J.\ Phys.\ C \textbf{14}, R415 (2002); Phys.\ Scr.\ \textbf{102}, 69 (2002).

\bibitem{Pittaway:2021yby}
I.~B.~Pittaway and F.~G.~Scholtz,
J.\ Phys.\ A \textbf{56}, 165303 (2023).

\bibitem{Nandi:2023tfq}
P.~Nandi and F.~G.~Scholtz,
Annals Phys.\ \textbf{464}, 169643 (2024).

\bibitem{Nandi:2022noj}
P.~Nandi,
Ph.D.\ thesis, University of Calcutta, Department of Physics, India (2022),
\href{https://arxiv.org/abs/2209.04758}{arXiv:2209.04758}.

\bibitem{PhysRevD.103.044017}
S.~Kanno, J.~Soda, and J.~Tokuda,
Phys.\ Rev.\ D \textbf{103}, 044017 (2021).

\bibitem{4}
B.~S.~DeWitt, Phys.\ Rev.\ \textbf{162}, 1239 (1967).

\bibitem{5}
R.~Penrose, Gen.\ Relativ.\ Gravit.\ \textbf{28}, 581 (1996).

\bibitem{6}
L.~Smolin, \textit{Three Roads to Quantum Gravity} (Basic Books, New York, 2000).

\bibitem{7}
M.~B.~Green, J.~H.~Schwarz, and E.~Witten, \textit{Superstring Theory. Vol.~1: Introduction} (Cambridge University Press, Cambridge, England, 1988).

\bibitem{8}
C.~Rovelli and F.~Vidotto, \textit{Covariant Loop Quantum Gravity} (Cambridge University Press, Cambridge, England, 2014).

\bibitem{8+}
P.~Nandi, T.~Bhattacharyya, A.~S.~Majumdar, G.~Pleasance, and F.~Petruccione, arXiv:2503.13061 [hep-th].

\bibitem{9}
G.~'t~Hooft and M.~J.~G.~Veltman, Ann.\ Inst.\ Henri Poincaré A \textbf{20}, 69 (1974).

\bibitem{10}
M.~P.~Bronstein, Zh.\ Eksp.\ Teor.\ Fiz.\ \textbf{6}, 195 (1936).

\bibitem{11}
S.~Deser, Gen.\ Relativ.\ Gravit.\ \textbf{1}, 9 (1970).

\bibitem{11+}
M.~Arzano and F.~Kuipers, Phys.\ Rev.\ D \textbf{111}, 025010 (2025).

\bibitem{Breuer1999}
H.-P. Breuer, B. Kappler, and F. Petruccione, Phys.\ Rev.\ A \textbf{59}, 1633 (1999).


\bibitem{nel1}
E.~Nelson, Phys.\ Rev.\ \textbf{150}, 1079 (1966).

\bibitem{nel2}
E.~Nelson, \textit{Quantum Fluctuations} (Princeton University Press, Princeton, 1985).

\bibitem{nel3}
E.~Nelson, \textit{Dynamical Theories of Brownian Motion}, 2nd ed. (unpublished manuscript, 2001), available at \url{http://www.math.princeton.edu/~nelson/books.html}.

\bibitem{gav}
B.~Gaveau, T.~Jacobson, M.~Kac, and L.~S.~Shulman, Phys.\ Rev.\ Lett.\ \textbf{53}, 419 (1984).

\bibitem{Przanowski:2024}
M.~Przanowski, J.~Tosiek, and Z.~Zawistowski,
Class.\ Quant.\ Grav.\ \textbf{41}, 095006 (2024).
%\href{https://doi.org/10.1088/1361-6382/ad2d6b}{doi:10.1088/1361-6382/ad2d6b}.

\bibitem{gu}
F.~Guerra and L.~M.~Morato, Phys.\ Rev.\ D \textbf{27}, 1774 (1983).

\bibitem{bohm}
D.~Bohm, Phys.\ Rev.\ \textbf{85}, 166 (1952); \textbf{85}, 180 (1952).

\bibitem{yas}
K.~Yasue, J.\ Math.\ Phys.\ \textbf{22}, 1010 (1981).

\bibitem{com}
G.~G.~Comisar, Phys.\ Rev.\ \textbf{138}, B1332 (1965).

\bibitem{wang}
M.~S.~Wang, Phys.\ Rev.\ A \textbf{37}, 1036 (1988).

\bibitem{gold}
S.~Goldstein, J.\ Stat.\ Phys.\ \textbf{47}, 645 (1987).

\bibitem{pav2}
M.~Pavon, J.\ Math.\ Phys.\ \textbf{40}, 5565 (1999).

\bibitem{pet}
N.~C.~Petroni and L.~M.~Morato, J.\ Phys.\ A \textbf{33}, 5833 (2000).

\bibitem{dunk}
J.~Dunkel and P.~H\"anggi, Phys.\ Rep.\ \textbf{471}, 1 (2009).

\bibitem{garb}
P.~Garbaczewski, Phys.\ Lett.\ A \textbf{164}, 6 (1992).

\bibitem{lind}
J.~Lindgren and J.~Liukkonen, Sci.\ Rep.\ \textbf{9}, 19984 (2019).

\bibitem{pap1}
L.~Papiez, J.\ Math.\ Phys.\ \textbf{23}, 1017 (1982).

\bibitem{yang}
C.-D.~Yang and S.-Y.~Han, Entropy \textbf{23}, 210 (2021).

\bibitem{yor}
V.~Yordanov, Sci.\ Rep.\ \textbf{14}, 6507 (2024).

\bibitem{biy}
I.~Bialynicki-Birula, Prog.\ Opt.\ \textbf{36}, 245 (1996).

\bibitem{weyl}
H.~Weyl, Z.\ Phys.\ \textbf{56}, 330 (1929).

\bibitem{bar}
J.~A.~Barandes, arXiv:1911.02515 [hep-th].

\bibitem{sud}
E.~C.~G.~Sudarshan, Phys.\ Rev.\ Lett.\ \textbf{10}, 277 (1963).

\bibitem{sp}
R.~J.~C.~Spreeuw, Found.\ Phys.\ \textbf{28}, 361 (1998).

\bibitem{gh}
P.~Ghose and A.~Mukherjee, Rev.\ Theor.\ Sci.\ \textbf{2}, 1 (2014).

\bibitem{q}
X.~Qian, B.~Little, J.~C.~Howell, and J.~H.~Eberly, Optica \textbf{2}, 611 (2015).

\bibitem{mar}
T.~W.~Marshall and E.~Santos, Phys.\ Rev.\ A \textbf{39}, 6271 (1989).

\bibitem{boy}
T.~H.~Boyer, Phys.\ Rev.\ D \textbf{11}, 790 (1975).

\bibitem{pen}
L.~de~la~Pe\~{n}a, A.~M.~Cetto, and A.~Vald\'es-Hern\'andez, \textit{The Emerging Quantum: The Physics Behind Quantum Mechanics} (Springer, New York, 2015).

\bibitem{pn}
J.~Applequist, J.\ Phys.\ A \textbf{22}, 4303 (1989).

\bibitem{pn2}
J.~Ramos, Gen.\ Relativ.\ Gravit.\ \textbf{38}, 773 (2006).

\bibitem{pn3}
J.~Ramos and R.~Gilmore, Int.\ J.\ Mod.\ Phys.\ D \textbf{15}, 505 (2006).

\bibitem{pn1}
J.~Ramos, Ph.D. thesis, Drexel University (2006), \url{http://hdl.handle.net/1860/1123}.

\bibitem{Pal:2021}
P.~B.~Pal,
``Covariant formulation of electrodynamics in isotropic media,''
\emph{Eur. J. Phys.} \textbf{43} (2021) 015204.

\bibitem{Nandi:prep}
P.~Nandi and P.~Ghose, ``in preparation.''

\bibitem{Breuer:2009}
H.-P.~Breuer, E.~Göklü, and C.~Lämmerzahl,
``Metric fluctuations and decoherence,''
\emph{Class. Quantum Grav.} \textbf{26} (2009) 105012.

\bibitem{Rovelli:2004tv}
C.~Rovelli,
\textit{Quantum Gravity},
Cambridge University Press (2004).
doi:10.1017/CBO9780511755804

\bibitem{Perez:2012wv}
A.~Perez,
``The Spin-Foam Approach to Quantum Gravity,''
\textit{Living Rev. Rel.} \textbf{16}, 3 (2013).
doi:10.12942/lrr-2013-3
[arXiv:1205.2019 [gr-qc]].






\end{thebibliography}
\end{document}